\documentstyle[12pt]{article}

\begin{document}

\newcommand{\beq}{\begin{equation}}
\newcommand{\eeq}{\end{equation}}

\begin{titlepage}

\begin{center}

\hfill  August 1995

\vskip .7in
 {\bf The Picard-Fuchs Equations, Monodromies and Instantons \\
 in the $N=2$ Susy Gauge Theories}

\vskip .3in

Shijong Ryang

\vskip .3in
{\em Department of Physics \\ Kyoto Prefectural University of Medicine \\
Taishogun, Kyoto 603 Japan}

\end{center}

\vskip .8in

\begin{center} {\bf Abstract} \end{center}
We construct the Picard-Fuchs equations of the $N=2$ supersymmetric
$SU(2)$ gauge theories with $N_f=0,1,3$  matter multiplets.  For the
$N_f=0$ theory from the solutions of the Picard-Fuchs equation the monodromy
matrices on the quantum moduli space are determined. We analyze the
Seiberg-Witten solutions to compute monodromies exactly and present the
instanton expansion  of the periods for the $N_f=0,3$ theories.

\vskip .3in

\end{titlepage}

Recently, Seiberg and Witten investigated the low energy properties of
$N=2$ supersymmetric $SU(2)$ gauge theories without or with matter
multiplets \cite{SW1,SW2}  and obtained the non-perturbative results such
as the exact expression for the metric on moduli space and the monopole
and dyon spectrums using a version of Olive-Montonen electric-magnetic
duality. The quntum moduli space of the gauge theory with no matter multiplet
was identified with the moduli space of torus and
further for the gauge theries with
$N_f$ matter multiplets the families of curves controlling the low energy
behavior were constructed. This approach was extended to the $SU(N_c)$
gauge theory with $N_f=0$ and the moduli space of
quantum vacua was shown to be provided
by that of a special set of genus $N_c-1$ hyperelliptic Riemann surfaces
\cite{KWTY}.  The generalization of it to include $N_f$ matter multiplets
 was performed and weak coupling monodromies were computed \cite{HO}.
The metric on moduli space was determined by a single holomorphic function,
whose structures are analogue of the special coordinates in the context
of Calabi-Yau moduli space.  The Picard-Fuchs (PF) equation indicating
the existence of a flat holomorphic connection on a certain  holomorphic
bundle has been found to be a second order equation for the
$N=2$\ \ $SU(2)$ gauge theory with no matter multiplet \cite{CDF}. With respect
 to the $SU(3)$  gauge theory with no matter multiplet
the PF equation has been solved and the leading instanton corrections in
the low energy quantum effective action have been studied \cite{KLT}.

In this letter by using the technique to transform a defining curve
into the Weierstrass form we construct the PF equations for $N=2$\ \ $SU(2)$
gauge theories without matter or with massless $N_f=1,3$ matter multiplets.
As for the $N_f=0$ theory the solutions of the PF equation will be related
with the Seiberg-Witten (SW) solutions. The monodromy matrices on the quantum
moduli space determined from the former solutions will be associated with
those determined from the latter solutions. By studying the properties of
the SW solutions for the $N_f=0$ as well as $N_f=3$ theories thoroughly
we will extract the non-perturbative instanton corrections to the metric
of the quantum moduli.

We start to consider the PF equation satisfied by the periods in the moduli
space of quantum vacua in $N=2$ supersymmetric gauge theories. So far the
PF equations were studied for the Landau-Ginzburg models whose marginal
deformations are associated with the complex structure moduli of the torus,
$K3$ and Calabi-Yau manifolds \cite{CF,KTS,R}. Here following the approach
 in Ref. \cite{KTS}
we write down the periods for a curve of torus in the Weierstrass form
$y^2=4x^3-g_2(u)x-g_3(u)$ as
\beq
\omega_i = \oint_{\gamma_i}\frac{dx}{y}\ ,
\eeq
where $\gamma_i$, $i=1,2$ are the two homology generators of the torus.
The PF equation is derived as
\beq
\frac{d^2\omega_i}{du^2} - \frac{a'_2}{a_2}\frac{d\omega_i}{ds} +
\left(\frac{a'_2a_1}{a_2} - a_3 \right)\omega_i = 0,
\label{omega}\eeq
where
\begin{eqnarray}
a_1= -\frac{\Delta'}{12\Delta}, \hspace{1cm}  a_2=\frac{3}{2\Delta}
(2g_2g'_3-3g'_2g_3),
\nonumber \\ a_3=\frac{1}{16\Delta}(g_2{g'}_{2}^{2}-12{g'}_{3}^{2}) + a'_1
\end{eqnarray}
with $\Delta=g_2^3-27g_3^2$\, which is the discriminant of the polynomial
$4x^3-g_2x-g_3$. The low energy quantum effective theory for the $N=2$
supersymmetric $SU(2)$ gauge theory has a single $N=2$ vector supermultiplet
which can be decomposed into a $N=1$ vector multiplet and a $N=1$ chiral
multiplet whose scalar component has the VEV denoted by $a$. The quantum
moduli space of it is given by the moduli space of the family of genus
one Riemann surface $y^2=(x-1)(x+1)(x-u)$ parametrized by the gauge
invariant quantity $u$ which becomes $a^2/2$ for large $u$. A change of
variables $x'=(x-u/3)/2$ , $y'=y/\sqrt2$ makes it's curve in the Weierstrass
form with $g_2=1+u^2/3$, $g_3=u^3/27-u/3$. From (\ref{omega}) we obtain
\beq
\frac{d^2\omega_i}{du^2} + \frac{2u}{u^2-1}\frac{d\omega_i}{du} +
\frac{\omega_i}{4(u^2-1)} = 0,
\label{domega}\eeq
which shows the $Z_2$ symmetry $u\rightarrow-u$ and contains two singular
points $u=1, -1$ which are associated with two massless states, a monopole
 and a dyon.  This PF equation is rewritten in terms of $z=u^2$ as
\beq
[(1-z)z\frac{d^2}{dz^2} + \left(\frac{1}{2}-\frac{3z}{2}\right)\frac{d}{dz}
- \frac{1}{16} ]  \omega_i = 0,
\eeq
whose two independent solutions around $z=0$, that is,  $u=0$ with suitable
 normalizations are expressed in terms of the hypergeometric function as
\begin{eqnarray}
f_1=\frac{\Gamma^2(\frac{1}{4})}{\Gamma(\frac{1}{2})}F\left(\frac{1}{4},
\frac{1}{4},\frac{1}{2},u^2\right), \nonumber \\
f_2=\frac{\Gamma^2(\frac{3}{4})}{\Gamma(\frac{3}{2})}uF\left(\frac{3}{4},
\frac{3}{4},\frac{3}{2},u^2\right),
\label{f}\end{eqnarray}
which indicate logarithmic singularities at $u=\pm1$. Each singularity at
$u=1$ and $u=-1$ leads to the monodromy matrices in the basis
$(f_1,f_2)^t$
\beq
\begin{array}{ccc} T_1=\left(\begin{array}{cc} 1-i & i \\ -i & 1+i
\end{array} \right) & , & T_{-1}=\left(\begin{array}{cc} 1-i & -i \\
i & 1+i \end{array} \right) \end{array}
\label{T}\eeq
around $u=1$ and $u=-1$ respectively, whose entries are complex numbers
\cite{R}.
The PF equation (\ref{domega}) is
in the same form as the Landau-Ginzburg model with  superpotential of
modality-one singularity type $X_9$ \cite{A}.
We turn to the $N=2$\ \ $SU(2)$ gauge theories with $N_f$ matter
multiplets\footnote{As for the $N_f>0$ theories we use the notation of
Ref.\cite{SW2}.}.
As for the $N_f=1$ theory with zero bare mass , whose curve is given by
$y^2=x^2(x-u)+{\Lambda_1}^6$ we make the same change of variables as
$N_f=0$ to have $g_2=u^2/3$, $g_3=u^3/27-{\Lambda_1}^6/2$. The PF equation
can be expressed in terms of $t=(4/27)^{1/3}u/{\Lambda_1}^2$ as
\beq
\frac{d^2\omega_i}{dt^2} + \frac{2t^3+1}{t(t^3-1)}\frac{d\omega_i}{dt} +
\frac{t}{4(t^3-1)}\omega_i = 0,
\eeq
which shows $Z_3$ symmetry under  $t\rightarrow\exp(2in\pi/3)t$
$(n=0,1,2)$. It includes three singular points
$u=3{\Lambda_1}^2/4^{1/3}\exp(2in\pi/3)$  $(n=0,1,2)$\ permuted by the
$Z_3$ symmetry, which correspond to three massless states, a monopole
and two dyons whose magnetic and electric charges $(n_m,n_e)$ are given by
$(1,0)$, $(1,1)$, $(1,2)$. A change of variable $t^3=z$ yields
\beq
[(1-z)z\frac{d^2}{dz^2} + \frac{1-4z}{3} -\frac{1}{36}]\omega_i =0
\eeq
with two independent solutions $F(1/6,1/6,1/3,z)$, $z^{2/3}F(5/6,5/6,5/3,z)$,
which indicate logarithmic singularities at $z=1$ , that is,
$u=3{\Lambda_1}^2/4^{1/3}\exp(2in\pi/3)$ $(n=0,1,2)$. In the massless $N_f=3$
th
the associated curve is given by $y^2=x^2(x-u)-(x-u)^2$, which can be
transformed into the Weierstrass form with $g_2=(u^2-4u+1)/3$,
$g_3=(2u^3+15u^2-12u+2)/54$ through a change of variables $x'=(x-(u+1)/3)/2$,
$y'=y/\sqrt2$. The PF equation is obtained in a compact form as
\beq
\frac{d^2\omega_i}{du^2} + \frac{8u-1}{u(4u-1)}\frac{d\omega_i}{du}
+ \frac{1}{u(4u-1)}\omega_i =0,
\eeq
which has no global symmetry. The regular singularities at $u=0,1/4$
correspond to two massless states with charges $(n_m,n_e) =(1,0), (2,1)$.
The two independent solutions are provided by $F(1/2,1/2,1,4u)$,
$F_1(1/2,1/2,1,4u) + F(1/2,1/2,1,4u)\log4u$, where
$F_1(1/2,1/2,1,z)=2\sum_{n=1}^{\infty}((1/2)_n/n!)^2\sum_{r=0}^{n-1}
(2/(2r+1) - 1/(r+1))z^n$ with $(j)_n = \Gamma(j+n)/\Gamma(j)$. They include
    logarithmic singularities at
$u=0,1/4$. In this way we have observed that the PF equations of the
$N_f=0,1,3$ theories are obtained as the hypergeometric differential
ones with appropriate singularities and symmetries, and the solutions for
the $N_f=3$  theory show a structure quite different
from those for the $N_f=0,1$ theories.

{}From now on we shall return to the $N_f=0$ theory. The monodromy matrix
acts on the vector $(a_D, a)^t$ where $a_D$ is the magnetic dual of $a$.
The pair $(a_D,a)$ is considered as a holomorphic section of an $SL(2,Z)$
bundle over the punctured complex $u$ plane, which is defined
by contour integrals
$a_D=\oint_{\gamma_1}\lambda$, $a= \oint_{\gamma_2}\lambda$ with
$\lambda=dx\sqrt{2(x-u)}/2\pi\sqrt{x^2-1}$, where the one cycle $\gamma_1$
loops around the branch points $x=u,1$ and the other cycle $\gamma_2$
around $x=1,-1$ . Explicitly, these are expressed by
$a=\sqrt{2(u+1)}F(-1/2,1/2,1,2/(u+1))$ with regular behavior at $u=\infty$
and $a_D=i(u-1)/2\:F(1/2,1/2,2,(1-u)/2) = i(u-1)/\sqrt{2(u+1)}\:F(1/2,3/2,
2,(u-1)/(u+1))$ , which contains a logarithmic singularity at $u=\infty$
as shown by
\beq
a_D= \frac{i}{\pi}a\log\frac{u+1}{2} + \frac{i}{\pi}(u-1)
\sum_{n=0}^{\infty}\frac{(\frac{1}{2})_n(\frac{3}{2})_n}
{(n!)^2} k_n \left(\frac{2}{u+1}\right)^{n+1/2}
\label{aD}\eeq
with $k_n=2\psi(n+1)-\psi(n+1/2)-\psi(n+3/2)$ where $\psi$ is the digamma
function. This expression yields directly a transformation
$a_D\rightarrow -a_D +2a$ under a circuit of the $u$ plane at large $u$ ,
being accompanied with $a\rightarrow -a$ . This result confirms the
derivation in the weak coupling limit using  only the asymptotic
expression $a\approx \sqrt{2u}$,
$a_D\approx i\sqrt{2u}/\pi\:\log{u}$\cite{SW1}. The regular behavior of $a_D$
at $u=1$ is compared with the logarithmic singularity of
$a=\sqrt{(u+1)/2}(F(1/2,1/2,1,2/(u+1)) + F(-1/2,3/2,1,2/(u+1))$ at $u=1$.
It is described by $a\approx ia_D/\pi\:\log{((u-1)/(u+1))}$
 which is in the dual
form of (\ref{aD}). Therefore definitely we obtain
 $a_D\rightarrow a_D$, $a\rightarrow a-2a_D$ when $u$ loops around $1$.
 This confirmative derivation of the strong coupling monodromy
 is compared with the leading order prescription
$a_D\approx i/2\:(u-1)$, $a\approx 4/\pi-(u-1)/2\pi\:\log(u-1)$, where the
coefficient $(u-1)$ of the logarithm is identified with $a_D$ . Here we
focus our attention on the periods along the two nontrivial cycles
$\gamma_1 , \gamma_2$ , that is , the derivative of the electric and magnetic
coordinates $(a_D,a)$ with respect to $u$, which are expressed by
\beq
\frac{da}{du}=-\frac{\sqrt2}{4\pi}\oint_{\gamma_1}\frac{dx}{y}=
\frac{1}{\sqrt{2(u+1)}}F\left(\frac{1}{2},\frac{1}{2},1,\frac{2}{u+1}\right),
\label{au}\eeq
\beq
\frac{da_D}{du}=-\frac{\sqrt2}{4\pi}\oint_{\gamma_2}\frac{dx}{y} =
\frac{i}{2}F\left(\frac{1}{2},\frac{1}{2},1,\frac{1-u}{2}\right).
\label{aDu}\eeq
In deriving them we represent each contour integral in terms of the
hypergeometric function. Alternatively direct derivative of the above obtained
expressions of $(a_D,a)$ about $u$ together with a recursion relation of
the hypergeometric functions yields them. Our task is to associate them
with the  solutions of the PF equation.
Since $F(1/2,1/2,1,(1-u)/2)=(\Gamma(1/2)/\Gamma^2(3/4))\:F(1/4,1/4,1/2,u^2)
 +  (\Gamma(-1/2)/\Gamma^2(1/4))\:uF(3/4,3/4,3/2,u^2)$ and $a'$
is rewritten as
$1/\sqrt{2u}\:F(1/4,3/4,1,1/u^2)$, the basis vector $({a_D}',a')^t$ is
transformed into $(f_1,f_2)^t$ of (\ref{f}) as
\beq
\left(\begin{array}{c}{a_D}' \\ a' \end{array} \right) =
\frac{1}{4\pi} \left( \begin{array}{cc} i & -i \\ 1+i & 1-i \end{array} \right)
\left( \begin{array}{c} f_1 \\ f_2 \end{array} \right) ,
\eeq
which is denoted by $ \Pi = P \omega$ . Then we can derive the monodromy
matrices around $u=1$ and $u=-1$ as $M_1=\left(\begin{array}{cc} 1 & 0 \\
-2 & 1 \end{array} \right)$ , $M_{-1} = \left( \begin{array}{cc} -1 & 2 \\
 -2 & 3 \end{array} \right)$ through
  $M_1=PT_1P^{-1}$, $M_{-1}=PT_{-1}P^{-1}$ with
$T_1,T_{-1}$ in (\ref{T}). We note that the monodromy matrices in the basis
$({a_D}',a')$ are the same as in the basis $(a_D,a)$ . It should be emphasized
that the quantum moduli space of $N=2$\ \ $SU(2)$ gauge theory is related with
the moduli space of complex deformation of the torus , in particular, of type
 $X_9$. From ${a_D}'=i/4\pi\:(f_1-f_2)$ and $F(j,j,2j,u^2)=(1-u^2)^{-j}
F(j,j,2j,u^2/(u^2-1))$ with $j=1/4, 3/4$ ,which contains a logarithmic
singularity at $u=\infty$, we obtain
\beq
\frac{da_D}{du}=\frac{i}{2\pi}\frac{da}{du}\log(1-u^2) + \frac{i}{4\pi}
\left(G(\frac{1}{4})-uG(\frac{3}{4})\right),
\label{aDG}\eeq
where $G(j)=\sum_{n=0}^{\infty}2((j)_{n}/n!)^2(\psi(n+1)-\psi(n+j))
(1-u^2)^{-(n+j)}$. This expression , which is compared
 with (\ref{aD}), and $a'=1/\sqrt{2u}\:
\sum_{n=0}^{\infty}(\frac{1}{4})_n(\frac{3}{4})_n\frac{1}{(n!)^2}u^{-2n}$
have the instanton contributions $u^{-2n}\; (n>0)$ from terms with an
odd number of instantons as well as an even number and reproduce the
monodromy matrix $M_{\infty}=\left(\begin{array}{cc} -1 & 2 \\ 0 & -1
\end{array} \right)$ 
about infinity in the $u$ plane. This structure is in contrast  to
the following expression derived from (\ref{aDu})
\begin{eqnarray}
\frac{da_D}{du}&=&\frac{i}{2}\sqrt{\frac{2}{u+1}}F\left(\frac{1}{2},\frac{1}{2}
,1,\frac{u-1}{u+1} \right) \nonumber \\
               &=&\frac{i}{\pi}\frac{da}{du}\log{\frac{u+1}{2}} +
    \frac{i}{2\pi}\sum_{n=0}^{\infty}\left(\frac{(\frac{1}{2})_n}{n!}
    \right)^2 k_n \left(\frac{2}{u+1}\right)^{n+1/2},
\label{aDul}\end{eqnarray}
where $k_n=2(\psi(n+1)-\psi(n+1/2)=2(\sum_{r=1}^{n}(1/r-2/(2r-1))+\log4)$.

Moreover, we consider the massless $N_f=3$ curve $y^2=(x-u)(x^2-x+u)$,whose
right-hand side has zeroes at $x=u$ and $x=x_{\pm}=(1{\pm}i\sqrt{4u-1})/2$
 for the
region $u>1/4$ which includes the semiclassical regime of $u\rightarrow
\infty$ . The periods are defined by $a'=\oint_{\gamma_1}\omega$, ${a'}_D
=\oint_{\gamma_2}\omega$ with $\omega=(\sqrt2/8\pi)dx/y$. The contours
$\gamma_1$, $\gamma_2$ loop around the pair points $(x_{+},x_{-})$,
$(x_{+},u)$ respectively. Contrary to the $N_f=1$ theory we can get
analytic expressions for $a',{a'}_D$
\beq
\frac{da}{du}=\frac{1}{2(2u-1+i\sqrt{4u-1})^{1/2}}F\left(\frac{1}{2},
\frac{1}{2},1,\frac{2}{1-iX}\right) ,
\label{auX}\eeq
\beq
\frac{da_D}{du}=\frac{\sqrt{2}i}{4(i\sqrt{4u-1})^{1/2}}F\left(\frac{1}{2},
\frac{1}{2},1,\frac{1+iX}{2}\right)
\label{aDuX}\eeq
with $X=(2u-1)/\sqrt{4u-1}$.
The above expressions have some similarity to (\ref{au}), (\ref{aDu})
 only differing
in $u$ replaced by $-iX$. The period $da/du$ can be shown to be
equated with $d\bar{a}/du=2^{-1}(2u-1-i\sqrt{4u-1})^{-1/2}
F(1/2,1/2,1,2/(1+iX))$ through a relation for the hypergeometric functions
and further be transformed into a manifestly real expression
\beq
\frac{da}{du}=\frac{1}{2\sqrt{2u-1}}F\left(\frac{1}{4},\frac{3}{4},1,
-\frac{1}{X^2}\right)=\frac{1}{2\sqrt{2u-1}}\sum_{n=0}^{\infty}
\frac{(\frac{1}{4})_n(\frac{3}{4})_n}{(n!)^2}\left(1-\frac{u}{1-\frac{1}{4u}}
\right)^{-n},
\label{auXr}\eeq
which shows the expected asymptotic behavior $1/2\sqrt{2u}$ for large $u$ in
the semiclassical region. In the general $N_f>0$ theory the section $(a,a_D)$
is expected to be
\begin{eqnarray}
a=\frac{1}{2}\sqrt{2u}\left(1+\sum_{n=1}^{\infty}a_{n}(N_f)\left(\frac
{\Lambda_{N_f}^{2}}{u}\right)^{n(4-N_f)}\right), \nonumber \\
a_D=i\frac{4-N_f}{2\pi}a\log{\frac{u}{\Lambda_{N_f}^2}} + \sqrt{u}
\sum_{n=0}^{\infty}a_{D_n}(N_f)
\left(\frac{\Lambda_{N_f}^2}{u}\right)^{n(4-N_f)}
\label{aaD}
\end{eqnarray}
with dynamically generated scale $\Lambda_{N_f}^2$, where  even number
instantons contribute \cite{SW2}. The $a$ in (\ref{aaD}) for $N_f=3$
 can be compared
with the above result (\ref{auXr}). On the other hand in order to make
(\ref{aDuX}) related with $a_D$ in (\ref{aaD}) we need to transform
 (\ref{aDuX}) into
\beq
\frac{da_D}{du}=\frac{\sqrt{2}i}{4(i\sqrt{4u-1})^{1/2}}
\left(\frac{\Gamma(\frac{1}{2})}
{\Gamma^2(\frac{3}{4})}F(\frac{1}{4},\frac{1}{4},\frac{1}{2},-X^2)-
iX\frac{\Gamma(-\frac{1}{2})}{\Gamma^2(\frac{1}{4})}F(\frac{3}{4},
\frac{3}{4},\frac{3}{2},-X^2)\right),
\label{aDu2}\eeq
whose hypergeometric functions are further changed into $(1\:+\:X^2)^{-1/4}
F(1/4,1/4,1/2,X^2/$\ $(X^2+1))$, $(1+X^2)^{-3/4}F(3/4,3/4,3/2,X^2/(X^2+1))$,
which include a logarithmic singularity at $u=\infty$. This logarithmic
singularity is expressed in a form similar to (\ref{aDG}) as
\beq
\frac{da_D}{du}=\frac{i}{2\pi}\frac{da}{du}\log{\left(
\frac{u}{1-\frac{1}{4u}}\right)} + \frac{1}{8\pi}\sqrt{\frac{i}{u}}
\left(H(\frac{1}{4})+i(1-\frac{1}{2u})H(\frac{3}{4})\right)
\label{aDH}\eeq
with $H(j)=\sum_{n=0}^{\infty}2((j)_n/n!)^2(\psi(n+1)-\psi(n+j))
(u/(1-1/4u))^{-n}$, whose behavior in the $u$ variable is the wanted one
compared with the instanton expansion of the $a_D$ in (\ref{aaD}) for $N_f=3$.
The relation (\ref{aDu2}) corresponds to ${a_D}'=i/4\pi(f_1-f_2)$ which has a
role to produce (\ref{aDG}). From (\ref{aDH}) together with
(\ref{auX}) the monodromy matrix
around $u=\infty$ can be exactly extracted as
$M_\infty=\left(\begin{array}{cc} -1 & 1 \\ 0 & -1 \end{array} \right)$.

Finaly we would like to present the alternative compact expression of
$a',{a_D}'$ as well as $a, a_D$ for the $N_f=0$ theory.
\begin{eqnarray}
 \frac{da}{du}=\frac{k}{\pi}K(k) &,& \frac{da_D}{du}=
\frac{ik}{\pi}K'(k),  \nonumber \\ a=\frac{4}{{\pi}k}E(k) & , &
a_D=\frac{4i}{{\pi}k}(K'(k)-E'(k)) ,
\end{eqnarray}
where $k=\sqrt{2/(u+1)}$ and $K(k)$, $E(k)$ are complete elliptic integrals
of the first and  second kinds. There is an interesting identity
$EK' + E'K - KK' = \pi/2$ , which turns out to be $ada_D/du-a_Dda/du=
2i/\pi$. The usual modular parameter $\tau_0=iK'/K$ of the torus
straightfowardly becomes ${a_D}'/a'$ which is the effective couplings
$\tau=\theta/2\pi + i4\pi/g^2$ combined by the dimensionless gauge coupling
constant and the theta parameter in the quantum vacuum . Through the expansion
of $q^{1/4}$ in powers of $1/(u+1)$ for $q=\exp{(i\pi\tau_0)}$, the metric on
the $u$ plane given by $(ds)^2 =$Im$\tau|da|^2$ is expressed in terms of
$u$ as
\beq
\mbox{Im}\tau=-\frac{4}{\pi}\log{\left((\frac{k}{4})^{1/2}(1+2(\frac{k}{4})^2
+15(\frac{k}{4})^4 + \ldots ) \right)},
\eeq
which resembles (\ref{aDul}). Conversely it is possible to describe $u$ in
terms
ratio ${a'}_D/a'$ by using $k=(\theta_2(0)/\theta_3(0))^2$ as
\beq
u=2\left(\frac{1+2\sum_{n=1}^{\infty}q^{n^2}}{2\sum_{n=0}^{\infty}
q^{(n+1/2)^2}}\right)^4 -1 .
\eeq

In conclusion we have derived the monodromy matrices definitely for the
$N=2$\ \  $SU(2)$ gauge theory with $N_f=0$ by taking advantage of the
interrelations between the hypergeometric functions. Our exact derivation
 provides support for the leading order prescription using the asymptotic
expression. They have been also obtained by studying
the solutions of the PF equation. Concerning the $N_f=1,3$ theories the
obtained second order differential equations for the periods yield solutions
expressed in terms of the hypergeometric function and it's analogue.
By making appropriate transformations of the hypergeometric functions, which
is suggested from analyzing the relation between the explicit expression ot
the SW solution for the $N_f=0$ theory and the two independent solutions of
the PF equation , we have gotten the desired power of the moduli parameter
in the instanton expansion of the periods for the $N_f=0,3$ theories.
Unlike the $N_f=0$ theory it is however unclear how the periods of the
$N_f=3$ theory are connected with the two independent solutions of the PF
equation. We hope the other representation of periods expressed in terms of
the complete elliptic integrals for the $N_f=0$ theory sheds new light on
further understanding of the quantum moduli space.

\end{document}